\documentstyle[12pt,a4,epsfig,a4wide]{article} 

\def\Jo#1#2#3#4{{#1} {\bf #2}, #3 (#4)}

\def\NPB{{Nucl. Phys.} {\bf B}}

\def\PLB{{Phys. Lett.}  {\bf B}}
\def\PRL{Phys. Rev. Lett.}
\def\PRD{{Phys. Rev.} {\bf D}}

\def\PRP{{Phys. Rep. }}
\def\EPC{{Eur. Phys. J.} {\bf C}}
\def\ZPC{{Z. Phys.} C}

\def\PTPS{Prog. Theo. Phys. Suppl.}

\def\st{\scriptstyle}

\def\ra{\rightarrow}
\def\be{\begin{equation}}
\def\ee{\end{equation}}

\def\gs{\mathrel{
   \rlap{\raise 0.511ex \hbox{$>$}}{\lower 0.511ex \hbox{$\sim$}}}}
\def\ls{\mathrel{
   \rlap{\raise 0.511ex \hbox{$<$}}{\lower 0.511ex \hbox{$\sim$}}}}
\newcommand{\obb}{0\mbox{$\nu\beta\beta$}}
\newcommand{\onbb}{neutrinoless double beta decay}

\newcommand{\ba}{\begin{array}{c}}
\newcommand{\baz}{\begin{array}{cc}}
\newcommand{\bad}{\begin{array}{ccc}}
\newcommand{\bea}{\begin{equation} \begin{array}{c}}
\newcommand{\eea}{ \end{array} \end{equation}}
\newcommand{\ea}{\end{array}}
\newcommand{\D}{\displaystyle}

\newcommand{\en}{\mbox{$\nu_e $}}
\newcommand{\mun}{\mbox{$\nu_{\mu}$ }}

\newcommand{\aen}{\mbox{$\overline{\nu}_e$ }}
\newcommand{\amun}{\mbox{$\overline{\nu}_{\mu} $}}

\newcommand{\mab}{\mbox{$\langle m_{\alpha \beta} \rangle $}}

\newcommand{\mmm}{\mbox{$\langle m_{\mu \mu} \rangle$}}
\newcommand{\mtta}{\mbox{$\langle m_{\tau \tau} \rangle $}}
\newcommand{\mmt}{\mbox{$\langle m_{\mu \tau} \rangle $}}
\newcommand{\mee}{\mbox{$\langle m_{ee} \rangle $}}
\newcommand{\met}{\mbox{$\langle m_{e \tau} \rangle $}}
\newcommand{\meu}{\mbox{$\langle m_{e \mu} \rangle $}}
\newcommand{\imab}{\mbox{$\langle \frac{\D 1}{\D m_{\alpha \beta}} \rangle $ }}
\newcommand{\immm}{\mbox{$\langle \frac{\D 1}{\D m_{\mu \mu}} \rangle $ }}
\newcommand{\imtt}{\mbox{$\langle \frac{\D 1}{\D m_{\tau \tau}} \rangle $ }}
\newcommand{\immt}{\mbox{$\langle \frac{\D 1}{\D m_{\mu \tau}} \rangle $ }}
\newcommand{\imee}{\mbox{$\langle \frac{\D 1}{\D m_{e e}} \rangle $ }}

\newcommand{\imem}{\mbox{$\langle \frac{\D 1}{\D m_{e \mu}} \rangle $ }}
\begin{document}
\newpage
\title{
\hfill { \bf {\small DO--TH 00/16}}\\
\hfill { \bf {\small hep-ph/0011050}}\\ \vskip 1cm  
\bf 
``Neutrinoless Double Beta Decay'' at a Neutrino Factory}
\author{W. Rodejohann$^a$\footnote{Email address:
rodejoha@xena.physik.uni-dortmund.de}$\,$ , $\, $  
K. Zuber$^b$\footnote{Email address:
zuber@physik.uni-dortmund.de}\\ \\
{\it $^a$Lehrstuhl f\"ur Theoretische Physik III,}\\
{\it $^b$Lehrstuhl f\"ur Experimentelle Physik IV,}\\ 
{\it Universit\"at Dortmund, Otto--Hahn Str.4,}\\
{\it 44221 Dortmund, Germany}}
\date{}
\maketitle
\thispagestyle{empty}
\begin{abstract}
We examine the prospects of detecting an  
analogous process of neutrinoless double beta decay at a 
neutrino factory from a high energy muon storage ring.  
Limits from LEP experiments, neutrinoless double beta decay as well 
as from global fits have to be 
incorporated and severely restrict the results. 
We investigate what limits on light and heavy effective Majorana 
neutrino masses 
can be obtained and compare them with existing ones. 
Discussed are also contributions from right--handed neutrinos and 
purely right--handed interactions. Other ``new physics'' 
contributions to the same final state 
might produce large event numbers. 
 
\end{abstract}

\newpage

\section{\label{eins}Introduction}
\begin{figure}[hb]
\epsfig{file=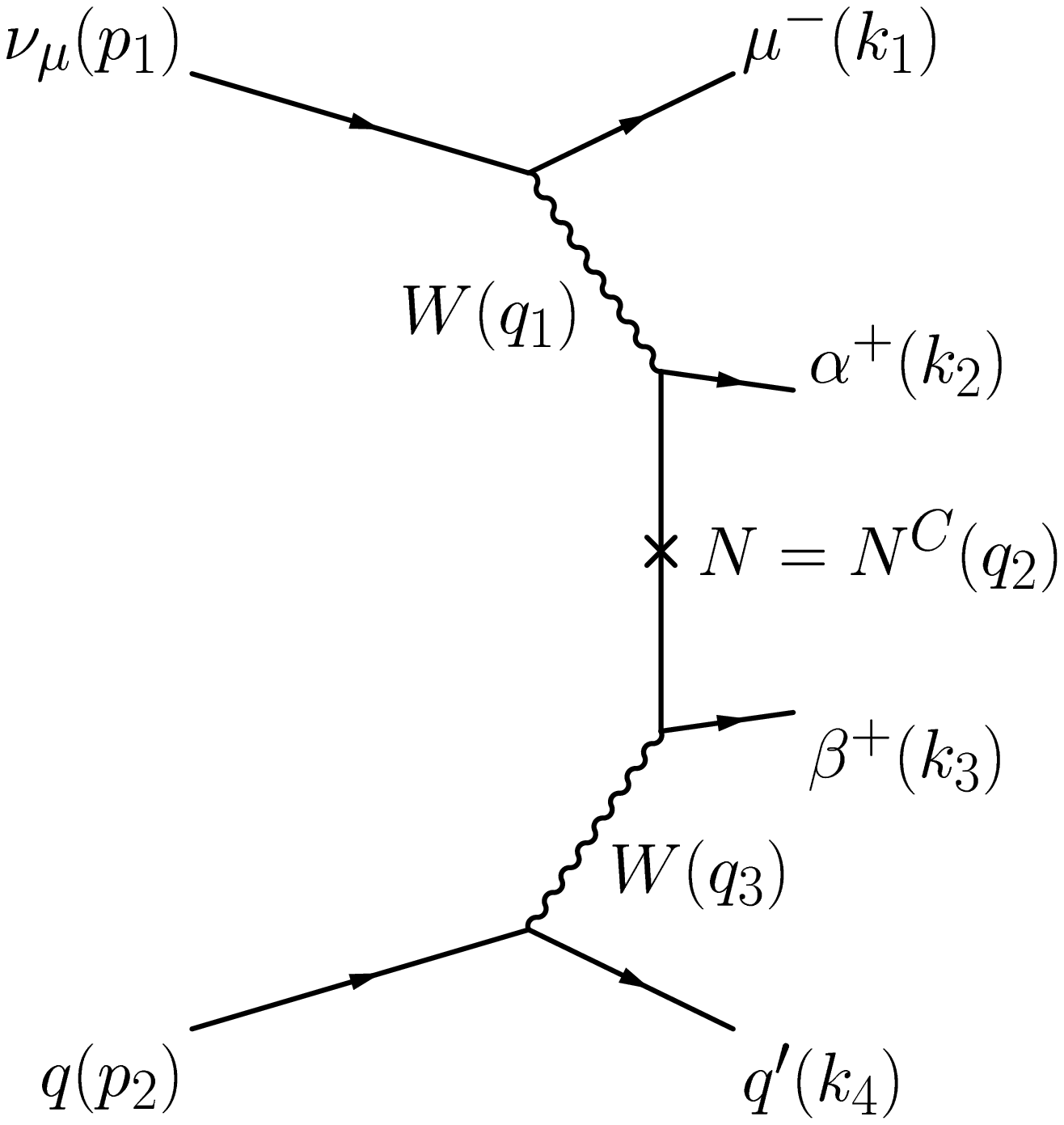,width=14cm,height=20cm}
\vspace{-6cm}
\caption{\label{diagramm}Diagram for
$\nu_\mu  q \rightarrow \mu^- \alpha^+ \beta^+ q'$.
Note that there is a crossed term and for $\alpha \neq \beta$ there are
two possibilities for the leptons to be emitted from. The leptonic part
also can be replaced by the corresponding other neutrino species.}
\end{figure}

\begin{figure}[ht]
\setlength{\unitlength}{1cm}
\vspace{-1cm}
\hspace{-1.4cm}
\epsfig{file=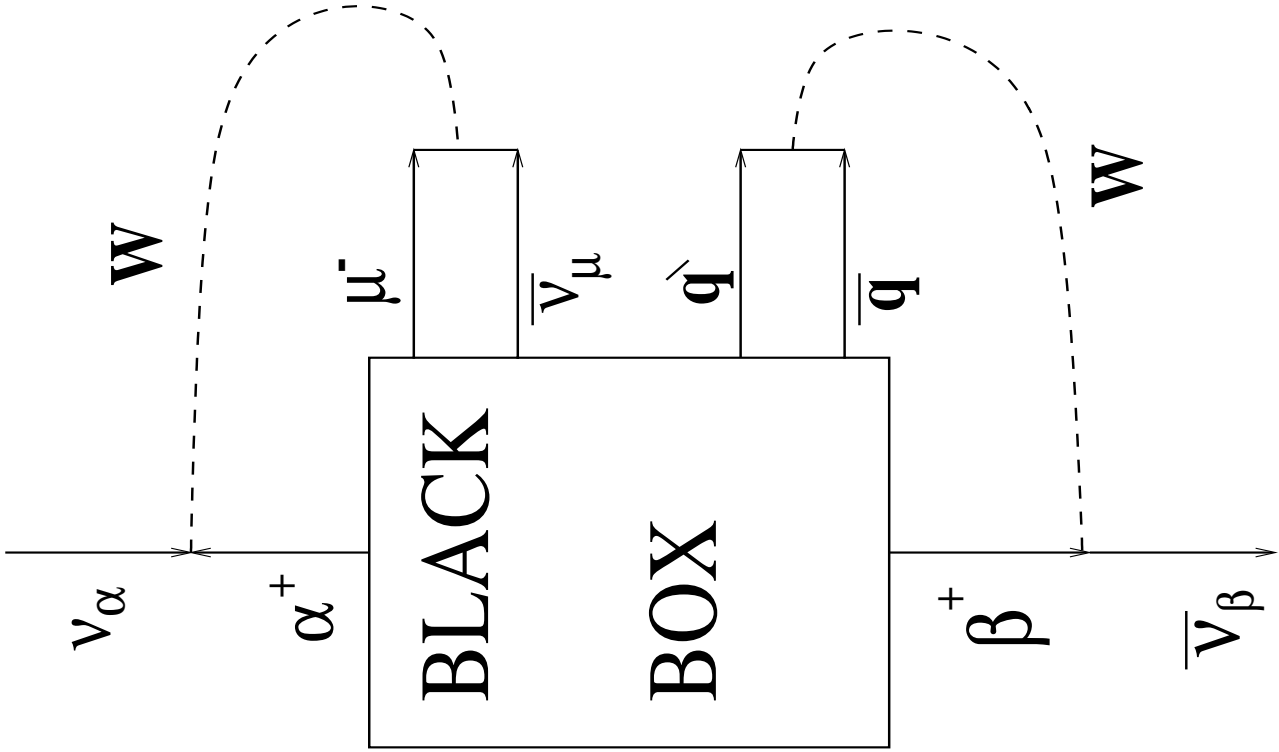,width=17cm,height=10cm}
\caption{\label{blackbox}Connection between Majorana mass term of
$\nu_\alpha$ and $\nu_\beta$ and the existence of process $\nu_\mu N \ra
\mu^- \alpha^+\beta^+ X$.}
\end{figure}

The physics potential of a muon storage ring is rich and exciting. 
Especially the option of using the neutrinos from the 
$\mu$ decay gathered much attention \cite{geer}. 
Typically the main focus lies in long baseline oscillation 
experiments \cite{LBL} 
with source--detector distances from 730 up to 
10000 km. This development is driven by the urge to find out about 
oscillation phenomena in more detail and to 
gain additional information, be it about CP violation, the sign of 
$\Delta m^2$, the size of $|U_{e3}|$ or 
the existence of sterile neutrinos.\\
An additional option 
is the usage of a detector directly at the storage ring site. 
Neutrino interactions of up to $10^{13}$/yr provide the possibility 
of high precision experiments regarding CKM matrix elements, 
structure functions, electroweak parameters, 
charm physics or other phenomena, see \cite{king} for some possibilities. 
As in any other new experiment, 
new physics may lurk in the results. In the light of recent  
developments in oscillation experiments, effects of massive neutrinos are 
hot candidates. Evidence for massive neutrinos 
and therefore physics beyond the Standard Model (SM) 
comes from the up--down asymmetry of atmospheric 
muon--neutrinos, the deficit of solar neutrinos 
and the LSND experiment. 
See \cite{kaireport} 
for more complete surveys. 
For example, the see--saw mechanism \cite{seesaw} might 
connect the very light known neutrinos to heavy neutrinos, 
which are usually assumed to be of Majorana nature. 
Majorana particles can show their presence not only by being directly 
produced, but also via indirect effects stemming from 
their $B-L$ violating mass term. The 
best known example for such a process 
is neutrinoless double beta decay (\obb{}) \cite{doi}, 
which results in limits on the effective electron neutrino 
Majorana mass \mee{}. The complete 3 $\times$ 3 matrix 
of (light) effective Majorana masses is defined as 
\bea \label{mab}
\mab = 
|(U \, {\rm diag}(m_1 , m_2 , m_3) U^{\rm T})_{\alpha \beta}| 
\\[0.3cm]
= \left| \D \sum m_i U_{\alpha i} U_{\beta i} \right|
\le \D \sum m_i | U_{\alpha i} U_{\beta i}|
\mbox{ with }  \alpha, \beta = e , \, \mu  , \, \tau ,
\eea  
where the sum goes over the mass eigenstates $m_i$. 
Conversely, the ``inverse effective mass'' is defined as 
\bea \label{imab}
\imab = 
|(U \, {\rm diag}(\frac{\D 1}{\D m_1} , \frac{\D 1}{\D m_2} , 
\frac{\D 1}{\D m_3}, 
\ldots) U^{\rm T})_{\alpha \beta}| 
\\[0.3cm]
= \left| \D \sum \frac{\D 1}{\D m_i} U_{\alpha i} U_{\beta i} \right|
\le \D \sum \frac{\D 1}{\D m_i} | U_{\alpha i} U_{\beta i}|
\mbox{ with }  \alpha, \beta = e , \, \mu  , \, \tau . 
\eea 
The sum over $i$ is not the same in 
Eqs.\ (\ref{mab}) and (\ref{imab}): For \mab{} it goes over all ``light'' 
mass eigenstates and in \imab over all ``heavy'' states. 
The attribute ``light'' or ``heavy'' depends on the energy scale of the 
process one considers to obtain information about the respective element.  
Note that we indicated that the sum goes up to a number 
greater than 3, it is however 
also possible that only one additional very heavy neutrino exists. 
Apart from theoretical prejudices, a priori 
we do not know how many there are.\\ 
The latter matrix might seem somewhat artificial; its form comes from the 
fact that heavy Majorana neutrinos force cross sections or branching ratios 
(typically processes in analogy to \obb) into a 
mass$^{-2}$ behavior. 
The knowledge of the elements of the matrices is 
rather poor, of course with the exception of \mee{} and \imee$\!\!$. 
At a neutrino factory, the following processes (see Fig.\ \ref{diagramm}) 
can be used to gain information about the other elements: 
\be \label{process}
\stackrel{(-)}{\nu}_{\! l}
N \ra l^\mp \alpha^\pm \beta^\pm X, \mbox{ where } 
l = e,\mu \mbox{ and } \alpha, \beta = e, \mu, \tau . 
\ee
Here $X$ denotes the hadronic final state. 
Because of very general arguments \cite{BL,RZ1}, 
the pure observation of this process 
guarantees a $B-L$ violating Majorana mass and in connection with 
SUSY a $B-L$ violating sneutrino mass term.  
This connection is depicted in Fig.\ \ref{blackbox} for the non--SUSY case. 
The precise determination of the mass of the intermediate Majorana neutrino 
will be very difficult, however, even the demonstration 
of Majorana mass terms will be an exciting and important result, since 
different models predict different mass matrices. 
In some models \mee{} is zero and therefore the only direct information 
about the mass matrix might come from neutrino oscillations. 
This complicates the situation, since only mass squared differences 
are measured and the additional phases induced by the 
Majorana nature are unobservable. Other experiments or 
cosmological arguments give total mass scales but the precise matrix 
is highly nontrivial to find \cite{ich2}. 
Thus, information about entries in \mab{} is very important. 
Similar arguments hold for the existence of heavy Majorana neutrinos. 
Note that 
the see--saw formula predicts their mass $m_N$ to lie in the range 
\be \label{massrange}
m_N \simeq \frac{m_D^2}{m_\nu} \simeq 10^2 \ldots 10^{18} \rm \; GeV \, , 
\ee 
where $m_D$ is a charged lepton or quark mass (i.e.\ electron to top quark) 
and $m_\nu$ the mass of a light neutrino ($10^{-5} \ldots 1$ eV as indicated 
by oscillation experiments). It turns out that the highest cross section 
of process (\ref{process}) 
is obtained for the lower region of this mass range.\\
In addition, if a $B-L$ violating process is detected, it is helpful to 
know if the ``mildly extended'' (i.e.\ just additional 
Majorana neutrinos) 
SM can provide the signal or 
another theory, such as SUSY has to be considered.\\
The paper is organized as follows: 
In Section \ref{zwei} we discuss some properties of process 
(\ref{process}) and its application to neutrino factory kinematics. 
We review in Section \ref{drei} the status of direct experimental 
limits on \mab{}. For the first time we give --- using HERA data --- 
bounds on elements of \imab 
other than \imee and examine what new limits might be accomplished 
for different neutrino factories. It turns out that for 
muon energies higher than 500 GeV 
physically meaningful limits on \mab{} can be obtained. 
Then we summarize limits on heavy Majorana neutrinos and their mixing 
with SM particles. 
Regarding the prospects of detecting events from process (\ref{process}) 
we apply in Section \ref{vier} all these limits, which severely restricts 
the results.

\section{\label{zwei}The process and a neutrino factory}
\subsection{Kinematics}
\begin{figure}[hb]
\begin{center}
\epsfig{file=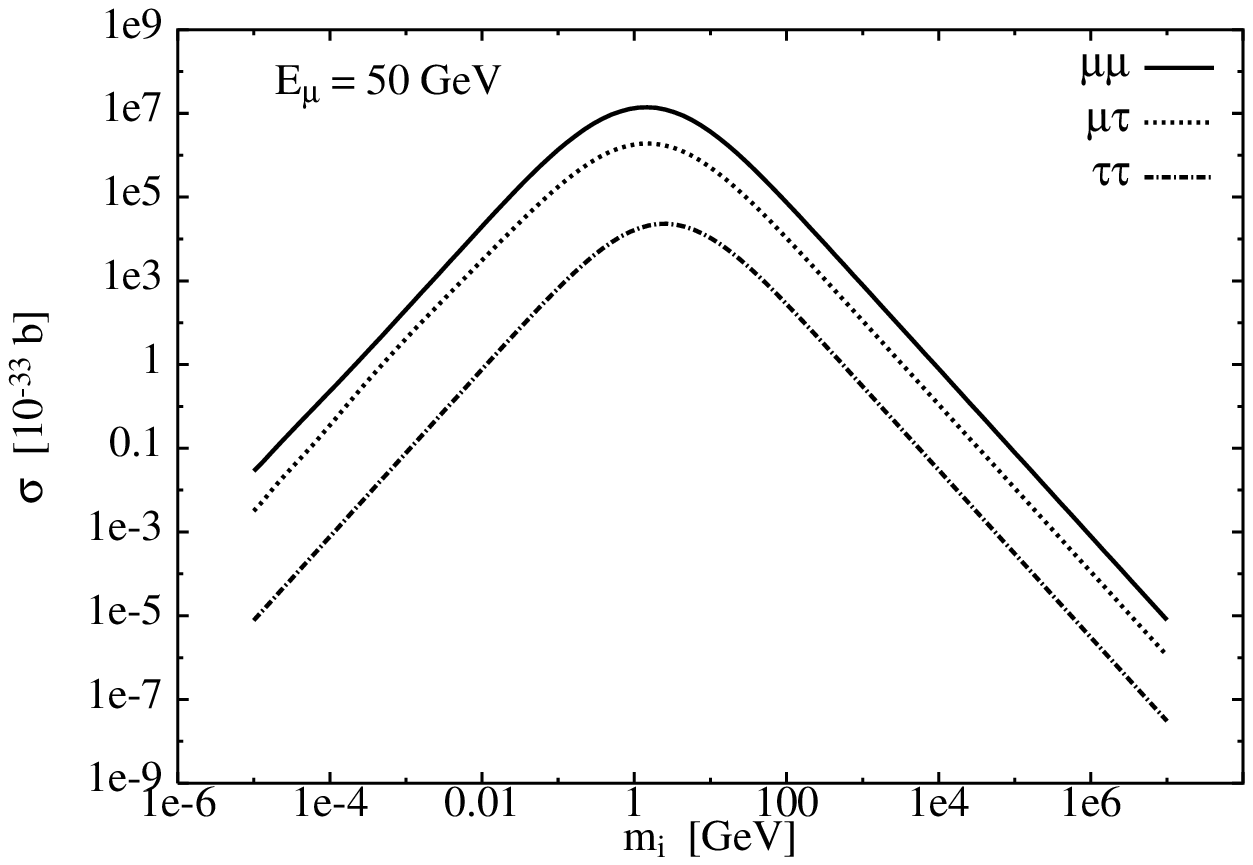,width=13cm,height=8cm}
\end{center}
\caption{\label{emu50}Total cross section for 
$\nu_\mu N \to \mu^- \alpha^+ \beta^+ X$ as a function of the Majorana mass 
for a $\mu^-$ energy of 50 GeV\@. No limits on U$_{\alpha i}$ are
applied.}
\end{figure}

\begin{figure}[ht]
\begin{center}
\epsfig{file=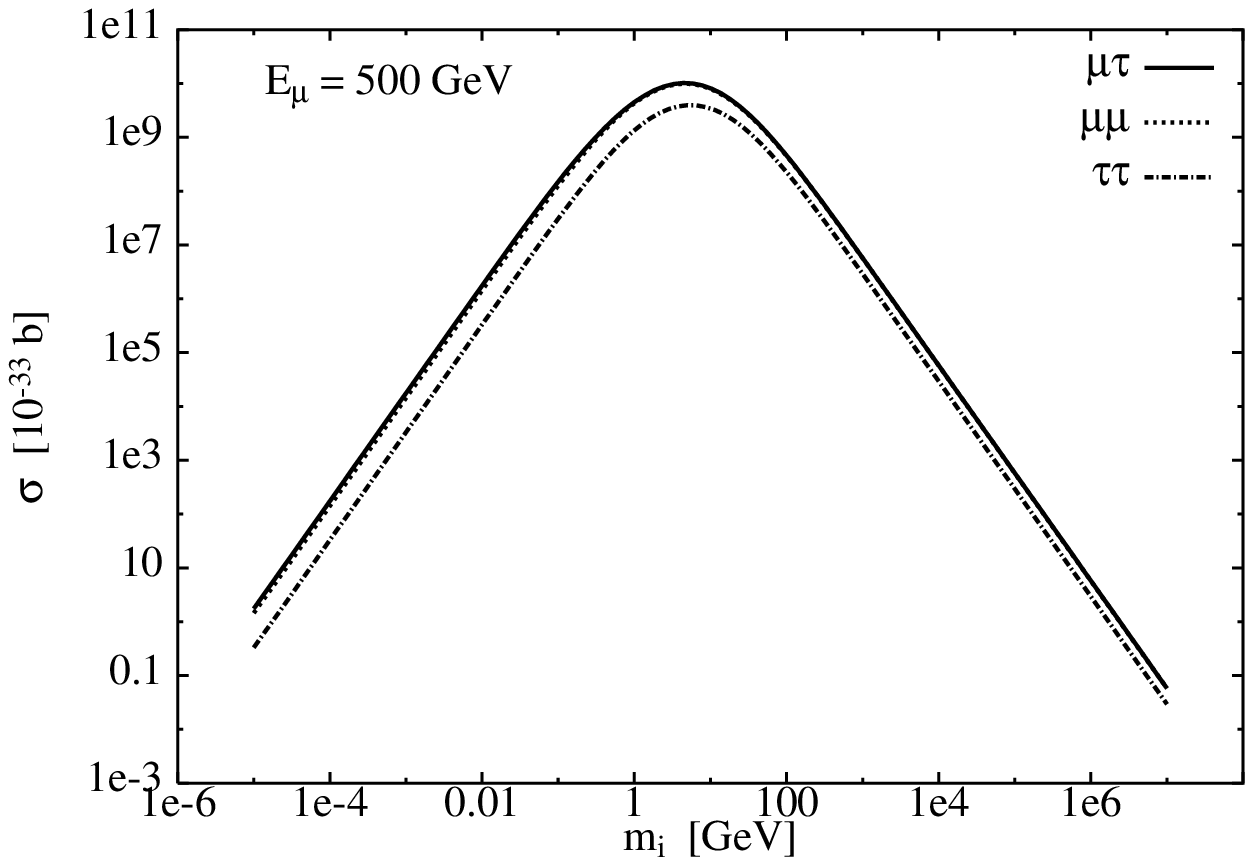,width=13cm,height=8cm}
\end{center}
\caption{\label{emu500}Total cross section for 
$\nu_\mu N \to \mu^- \alpha^+ \beta^+ X$ as a function of the Majorana mass 
for a $\mu^-$ energy of 500 GeV\@. No limits on U$_{\alpha i}$ are
applied.}
\end{figure}

\begin{figure}[hb]
\begin{center}
\epsfig{file=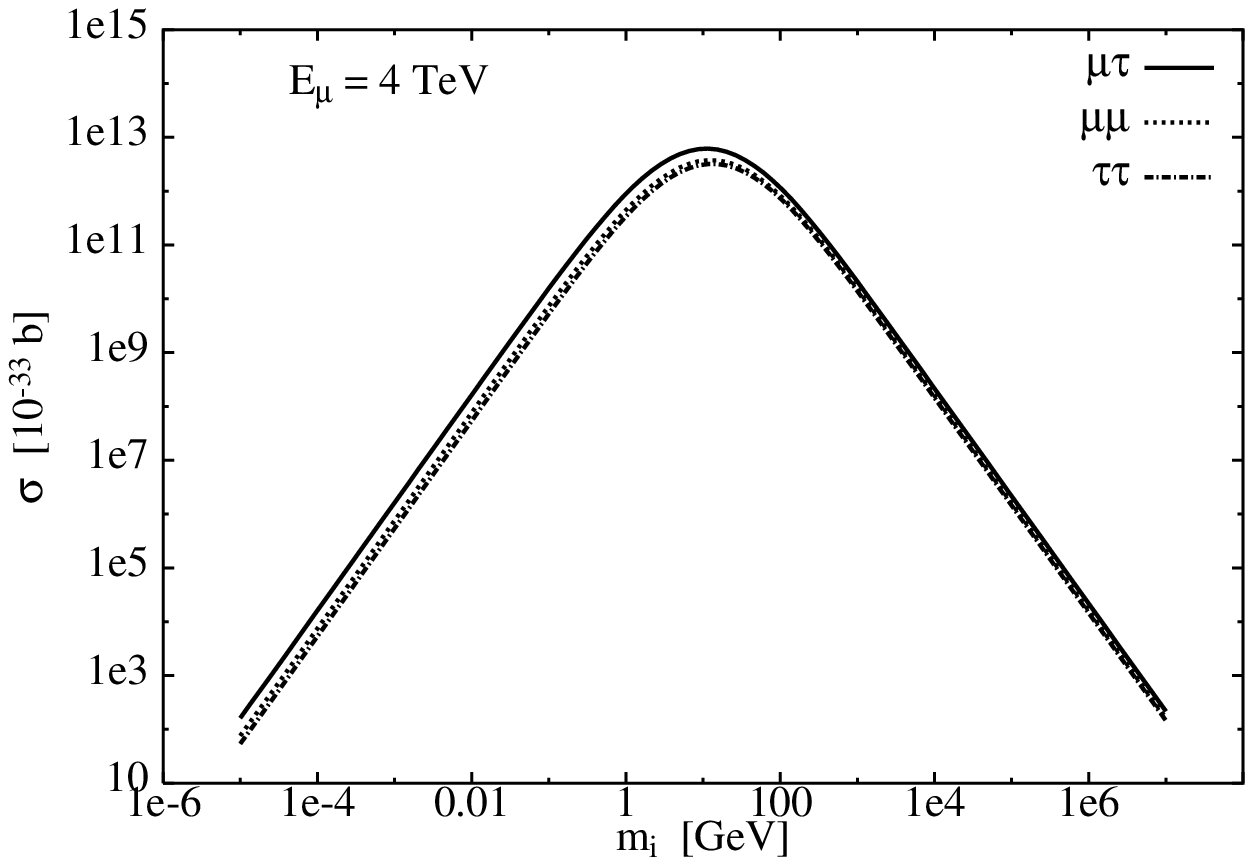,width=13cm,height=8cm}
\end{center}
\caption{\label{emu4}Total cross section for 
$\nu_\mu N \to \mu^- \alpha^+ \beta^+ X$ as a function of the Majorana mass 
for a $\mu^-$ energy of 4 TeV\@. No limits on U$_{\alpha i}$ are applied.}
\end{figure}

\begin{figure}[ht]
\begin{center}
\epsfig{file=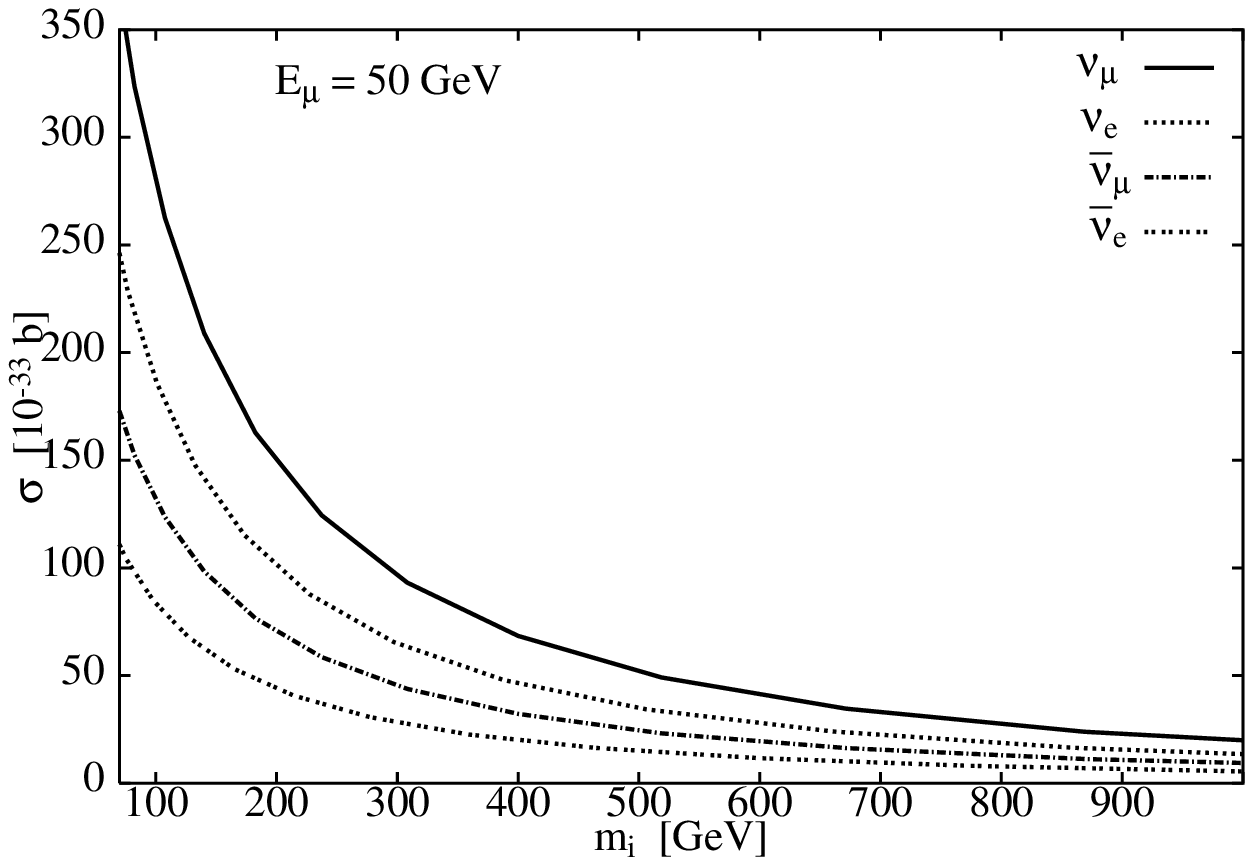,width=13cm,height=8cm}
\end{center}
\caption{\label{emu50tot}Total cross section for 
$\nu_l N \to l \mu \mu X$ for $\nu_\mu, \en, \amun$ and $\aen$ 
as a function of the Majorana mass 
for a $\mu$ energy of 50 GeV\@. The limits on U$_{\alpha i}$ are applied.}
\begin{center}
\epsfig{file=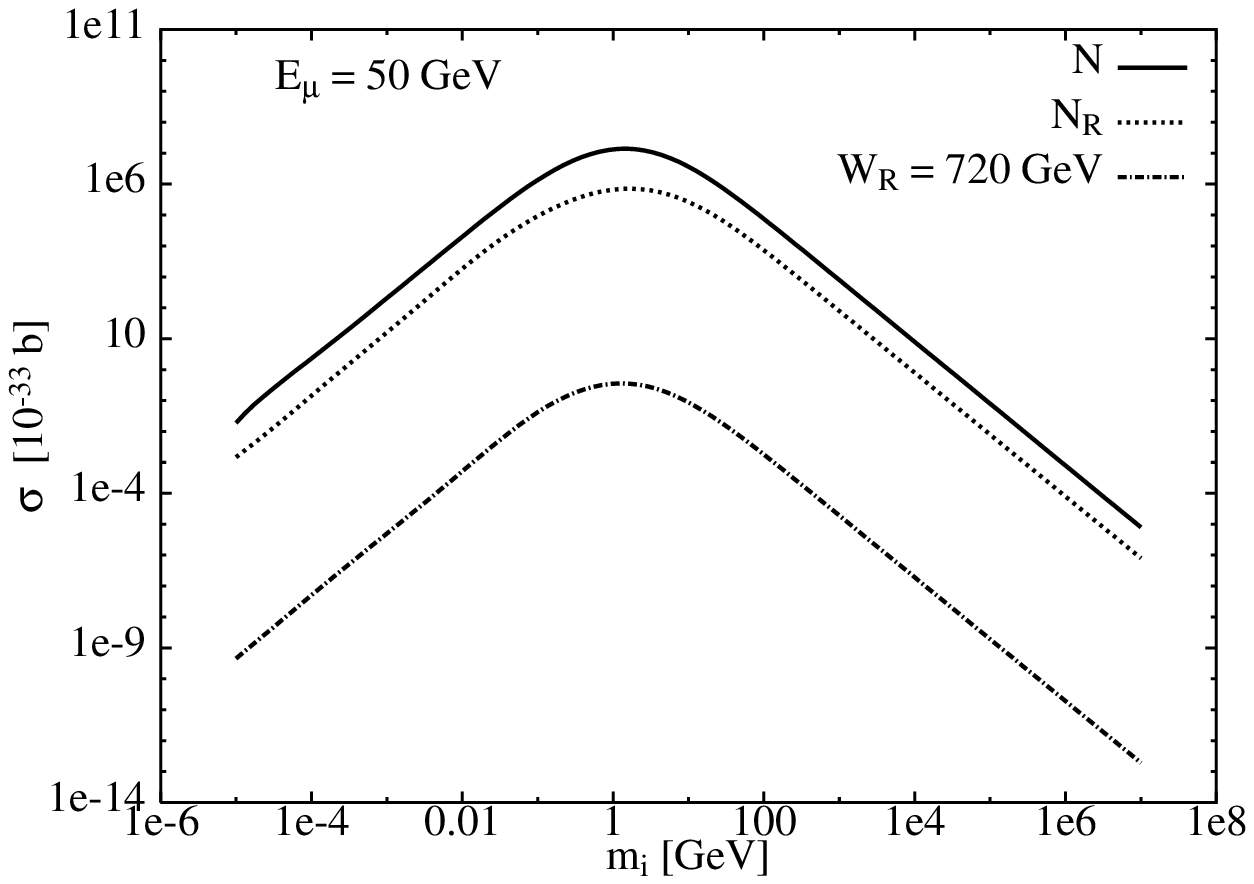,width=13cm,height=8cm}
\end{center}
\caption{\label{emu50R}Total cross section for 
$\nu_\mu N \to \mu^- \alpha^+ \beta^+ X$ as a function of the Majorana mass 
for a $\mu^-$ energy of 50 GeV and different possible realizations 
of the process. $N$ is a left--handed, $N_R$ a 
right--handed Majorana and $W_R$ denotes the process with a right--handed 
$W$ boson and a left--handed Majorana. 
No limits on U$_{\alpha i}$ are applied.}
\end{figure}

The Feynman diagram is shown in Fig.\ \ref{diagramm}, the calculation 
is straightforward and described in more detail in \cite{FRZ1}. 
For one eigenstate $m_i$ it was found: 
\bea \label{matrix}
|\overline{\mbox{$\cal{M}$}}|^2 (\nu_l  \, q \to l^- \, \alpha^+ \alpha^+ q') 
\equiv |\overline{\mbox{$\cal{M}$}_{--}}|^2 
= m_i^2 U_{\alpha i}^4 \, G_F^4 \, M_W^8 \,
2^{12} \, \frac{\D 1 }{\D (q_1^2 - M_W^2 )^2(q_3^2 - M_W^2)^2}
(p_1 \cdot p_2)  \\[0.4cm]
\left[ \frac{\D 1}{\D (q_2^2 - m_i^2)^2} (k_1 \cdot k_2)(k_3 \cdot k_4) +
\frac{\D 1}{\D (\tilde{q_2}^2 - m_i^2)^2}(k_1 \cdot k_3)(k_2 \cdot k_4)
\right. \\[0.4cm]
\left. - \frac{\D 1}{\D (q_2^2 - m_i^2)(\tilde{q_2}^2 - m_i^2)}
\mbox{\Large(}
(k_2 \cdot k_3)(k_1 \cdot k_4) - (k_1 \cdot k_2)(k_3 \cdot k_4)
- (k_1 \cdot k_3)(k_2 \cdot k_4) \mbox{\Large)} \right] . 
\eea     
Here $\tilde{q_2}$ denotes the momentum of the Majorana neutrino in the
crossed diagram, which has a relative sign due to the interchange
of two identical fermion lines.
In addition one has to include a factor $\frac{1}{2}$ to avoid
double counting in the phase space integration.           
Scattering with an antiquark and with an antineutrino is equivalent 
to the following simple replacements: 
\bea \label{repl}
|\overline{\mbox{$\cal{M}$}}|^2 
(\nu_l  \, \overline{q} \to l^- \, \alpha^+ \alpha^+ \overline{q'}) 
\equiv |\overline{\mbox{$\cal{M}$}_{-+}}|^2 = 
|\overline{\mbox{$\cal{M}$}_{--}}|^2 (p_2 \leftrightarrow k_4) , \\[0.4cm]
|\overline{\mbox{$\cal{M}$}}|^2 
(\overline{\nu}_l   \, q \to l^+  \, \alpha^- \alpha^- q') 
\equiv |\overline{\mbox{$\cal{M}$}_{+-}}|^2 = 
|\overline{\mbox{$\cal{M}$}_{-+}}|^2 , \\[0.4cm]
|\overline{\mbox{$\cal{M}$}}|^2 
(\overline{\nu}_l   \, \overline{q} \to l^+  \, \alpha^- \alpha^- 
\overline{q'})
\equiv |\overline{\mbox{$\cal{M}$}_{++}}|^2 =
|\overline{\mbox{$\cal{M}$}_{--}}|^2 .
\eea
Of course, the two leptons from the intermediate ``$WW \to \alpha \beta$'' 
diagram do not have to be of the same flavor: 
An interesting statistical effect \cite{RZ1} 
occurs when one considers the relative 
difference between, say, the $\mu\mu$ and the $\mu e$ final 
state (mass effects play no role for $e$ and $\mu$):
First, there is no phase space factor $\frac{1}{2}$ for the latter case. 
Then, there is the possibility that an electron is produced at the (``upper'')
$\nu l W$ vertex or at the (``lower'') $q q' W$ vertex. Both diagrams are
topologically distinct and thus have to be treated separately.
This means, four diagrams lead to the $\mu e$ final state, whereas only
two lead to the $\mu\mu$ final state.
We see that there is a relative factor 4 between the two cases.
Note though that now the interference terms are {\it added}
to the two squared amplitudes since there is no relative
sign between the two. This reduces the relative factor to about 3.\\ 
The details of our Monte Carlo program are given in \cite{FRZ1}, now an 
integration over the incoming neutrino energy spectrum has to be included. 
A simple phase space calculation for the $\mu \ra e \nu_e \nu_\mu$
decay gives for the normalized 
distribution in the lab frame: 
\bea \label{spectrum}
\frac{\D dN}{\D dE_\nu} (\nu_\mu) = \frac{\D 2}{\D E_\mu^3} E_\nu^2 
\left( 3 - 2 \frac{\D E_\nu}{\D E_\mu} \right) , \\[0.4cm]
\frac{\D dN}{\D dE_\nu} (\nu_e) = \frac{\D 12}{\D E_\mu^3} E_\nu^2 
\left( 1 - \frac{\D E_\nu}{\D E_\mu} \right) . 
\eea
The maximal neutrino energy is $E_\mu$ and the mean value is 
$\langle E_\nu \rangle = 7/10 \; (3/5) \, E_\mu$ for $\nu_\mu$ $(\nu_e)$.\\
In Figs.\ \ref{emu50} to \ref{emu4} we show the total 
cross section for the reaction 
$\nu_\mu N \to \mu^- \alpha^+ \beta^+ X$ with 
the $\mu\mu$, $\mu\tau$ and $\tau \tau$ final states for 
three different muon energies.  
For the above given statistical arguments to be valid, the mass 
of the final state leptons has to be negligible. This is the case for muon
energies higher than about 3 TeV, as can be seen from the figures, where the 
$\mu\tau$ final state is the leading signal for $E_\mu = $ 4 TeV\@. 
We set $U_{\alpha i} = 1$ in order to show the mass dependence 
of the signal; the slope of the curves is easily 
understood from the two extreme limits of 
\be
 \sigma \propto \frac{\D m_i^2}{\D (q^2 - m_i^2)^2}
\rightarrow \left\{ \baz m_i^2 & \mbox{ for } m_i^2 \ll q^2 \\[0.3cm]
                         m_i^{-2} & \mbox{ for } m_i^2 \gg q^2 \ea , \right.
\ee     
where $q$ is the momentum of the Majorana neutrino.\\
For a $\mu^- \mu^+$--collider or a muon storage ring four different 
signals are possible 
(corresponding to incoming $\nu_\mu$, \aen{}$\!\!$, \amun{} or \en{}), 
Fig.\ \ref{emu50tot} shows that the 
muon neutrino from the $\mu^-$ decay gives the highest cross section. 
In this figure we plotted the interesting area of the mass range 
given by the see--saw formula (\ref{massrange}) and applied also the 
limits on heavy neutrino mixing as explained in Section \ref{drei}. 
Finally, we give in Fig.\ \ref{emu50R} the cross section for 
$\nu_\mu N \to \mu^- \mu^+ \mu^+ X$ with two possible other realizations, 
namely via an intermediate right--handed Majorana neutrino and via 
right--handed interactions with a $W_R$--mass of 720 GeV, the 
current lower bound \cite{WR}. For the latter case, the matrix elements are 
identical whereas for the former one has to make the replacement 
($p_2 \leftrightarrow k_4 , \, p_1 \leftrightarrow k_1$) 
in Eqs.\ (\ref{matrix}) and (\ref{repl}).
It can be seen, that a left--handed heavy neutrino gives the highest 
contribution. Of course it is possible that all these realizations 
contribute and thus interfere.\\ 
We checked the dependence of the results on 
oscillation parameters by integrating over two--flavor formulas. 
Even for $E_\mu$ = 50 GeV, a detector--source distance of 1 km and 
LSND--like values of $\Delta m^2$ ($\simeq 0.1$ eV$^2$) 
and $\sin^2 2 \theta$ ($\simeq 10^{-3}$) the 
relative suppression of the signal was not more than $\cal{O}$($10^{-6}$). 
Inserting typical parameters of atmospheric or solar experiments has even 
less effect. 
\subsection{\label{paraset}Neutrino factories}
Several proposals for a muon storage ring have been discussed, the number 
of expected neutrino interactions differs. The formula used for the 
luminosity in units of cm$^{-2}$ s$^{-1}$ is \cite{mighty} 
\be \label{lumi}
\mbox{$\cal{L}$} = N_A f N_\mu l , 
\ee 
where $N_A$ is the Avogadro number, $N_\mu$ the number 
of muons injected in the ring per second, 
$f$ the fraction of the collider ring 
occupied by the production straight section and $l$ the mass depth 
of the target in g cm$^{-2}$. Typical numbers are $f$ = 0.02, 
$l = 1000$ g cm$^{-2}$ and $N_\mu = 10^{12} \ldots 10^{14}$ s$^{-1}$. Let us 
be optimistic and assume $N_\mu = 10^{14}$ s$^{-1}$ with a ``year'' of 
$10^{7}$ s running time. With this parameter set one gets 
$\mbox{$\cal{L}$}$ $\simeq 10^{39}$ cm$^{-2}$ s$^{-1}$. The neutrinos from 
the $\mu$--decay will all end inside the detector since their opening angle 
is just $\theta \simeq 1/\gamma_\mu = m_\mu/E_\mu$. Typical distances 
between detector and muon ring are $10^2$ to $10^3$ m, 
discussed energies go up to $10^6$ GeV\@. 
A complete scope of all possible options 
is not our aim. If definite plans for experiments and machines 
are made, our results 
can easily be rescaled with the help of relation (\ref{lumi}).

\section{\label{drei}Limits on neutrino parameters}
As expected, \obb{} provides us with the best limit of all entries in 
\mab{} and \imab$\!\!$. Recently, other elements of the 
mass matrix were investigated and for the first time limits 
on the $\tau$ sector 
of \mab{} were given \cite{FRZ2}. 
The process discussed was 
$e^+ p \to \aen{} \alpha^+ \beta^+ X$ at HERA and gave 
bounds on \met{}, \mmt{} and \mtta$\!$. 
In \cite{kaiplb} improved limits on \mmm{} (via $K^+ \to \pi^- \mu^+ \mu^+$) 
and \meu{} ($\mu^-$--$e^+$ conversion on titanium) are given. 
Together with the \obb{} limit \cite{obb} the current situation is 
as follows:
\be \label{meffres}
\mab  \! \ls
 \left( \bad 2 \cdot 10^{-10} \; \cite{obb} 
& 1.7 \; (8.2) \cdot 10^{-2} \; \cite{kaiplb} 
& 4.2 \cdot 10^{3} \; \cite{FRZ2} \\[0.2cm]
&  5.0 \cdot 10^{2} \; \cite{kaiplb} 
& 4.4 \cdot 10^{3} \; \cite{FRZ2} \\[0.2cm]
&  & 2.0 \cdot 10^{4} \; \cite{FRZ2} 
\ea \right) \rm GeV .
\ee
There is a spread over 14 orders of magnitude. For \meu{} two values 
are given, depending on the spin configuration of the final state protons. 
Note that for all entries except for the $ee$ element the limits 
lie in the unphysical region, e.g.\ for a $\met = 4.2 \cdot 10^{3}$ 
GeV the cross section is proportional to $m^{-2}$ and not to 
$m^2$ as assumed to get the limit. Improvement on most values might be 
expected from $B$ decays \cite{kaiplb,ich}.\\
For \imee a limit from \obb{} exists \cite{obbH}, for which a heavy 
neutrino has $m_i \gs 1$ GeV\@. 
Beside \onbb{} there are other ways 
to get information about heavy neutrinos: 
The LEP machine can produce heavy neutral leptons via 
$e^+ e^- \to N \overline{N}$, the most stringent limits come from 
the L3 collaboration \cite{L3}; they exclude masses below 70 to 80 GeV, 
depending on the charged lepton they couple to ($e$, $\mu$ or $\tau$). 
On the other hand, if heavy neutrinos mix with their light SM counterparts, 
they should alter the results for $\mu$ decay, $\nu$ scattering and so on. 
Global fits then limit the mixing parameters \cite{Ulimits}, in total 
the limits read: 
\be \label{neulim}
\baz
\sum |U_{ei}|^2 < 6.6 \cdot 10^{-3},  & m_i > 81.8 \; \rm GeV ,\\[0.3cm]
\sum |U_{\mu i}|^2 < 6.0 \cdot 10^{-3},  & m_i > 84.1 \; \rm GeV ,\\[0.3cm]
\sum |U_{\tau i}|^2 < 1.8 \cdot 10^{-2},  & m_i > 73.5 \; \rm GeV .
\ea 
\ee
In Section \ref{obs?} we will discuss how these bounds restrict the 
possibilities of observing the \obb{} analogue at a neutrino factory. 
Before that we apply the 
procedure from \cite{FRZ2} again to the HERA data and 
gain limits on the other elements of \imab$\!\!$. Here, heavy neutrinos 
must have $m_i \gs 100$ GeV\@. The matrix reads:  
\be \label{imeffres}
\imab \! \ls
 \left( \bad 
1.1 \cdot 10^{-8} & 5.4 \cdot 10^{-3} & 8.6 \cdot 10^{-3} \\[0.2cm]
                  & 8.4 \cdot 10^{-3} & 9.0 \cdot 10^{-3} \\[0.2cm]
                  &                   & 0.1    
\ea \right) \rm GeV^{-1} .
\ee
Now there is only a spread of 7 orders of magnitude. 
All non--$ee$ entries are unphysical, e.g.\ for the $\mu\mu$ element 
one gets with the bound from Eq.\ (\ref{neulim})
\be
m_i \gs \frac{\D |U_{\mu i}|^2}{\D  \immm} \simeq 0.7 \; \rm GeV . 
\ee
All limits on the same quantities for a {\it right--handed} Majorana 
neutrino with the usual couplings to the SM particles lie in the 
same order of magnitude. 
Table \ref{table1} shows what limits could be achieved for luminosity 
per year given by the parameter set after Eq.\ (\ref{lumi}). 
The improvement would be tremendous and already for 
muon energies higher than 500 GeV the bounds on \mab{} lie in the 
physical region: The limit on \mmm{} is about 3 GeV, where the slope of 
the cross section is still rising, i.e.\ proportional to 
$m_i^2$. 
For \imab{} the situation is different: For $E_\mu = 500$ GeV 
the limit on \imtt{} is about 0.2 GeV, which translates in $m_i \gs 0.1$ GeV, 
which is a light neutrino, i.e.\ in that region is 
$\sigma \propto m_i^2$. Here, energies around 10 TeV are required to get 
physical meaningful values.\\

\section{\label{vier}Detection of the process}
\subsection{Experimental considerations}
Because of the smallness of the cross section of the process
discussed here, one might ask if 
SM processes exist, which fake the signal. 
A discussion of this kind has already been done for 
trimuon production in $\nu N$ scattering at previous 
fixed target experiments, both experimentally \cite{trimuonexp} 
and theoretically \cite{trimuontheo}. 
This trimuon production has a ($-++$) signature. 
Due to the principle creation of conventional neutrino beams by using pion-- 
and kaon--decays, there is always a 
``$\amun$--pollution'' in the beam which can give a ($-++$) signal 
through muon pair production, be it radiatively or 
in the hadronic final state via e.g.\ vector meson production. 
These effects exist on the level of about $10^{-4}$ 
of the total observed
charged current events. 
Kinematical cuts to suppress this background, e.g.\ using 
the invariant mass or 
angular isolation have been developed. For previous experiments however, 
it was found \cite{FRZ1} that for trimuon production via Majorana 
neutrinos the signal--to--background ratio is far too small.  
However, for a muon storage ring we know exactly what neutrino flavor is 
coming in and thus in the case of $\mu^-$--decay there is no SM process 
to give a $(-++)$ event. 
The only exception is $\mun N \to \mu^- e^+ \mu^+ X $ which might 
be faked by a \aen CC event with $\mu^+ \mu^-$ production in 
the jet or via bremsstrahlung. However, as we will show below, final 
states with electrons cannot be expected due to the severe limits from 
\obb{}. Because of the $\nu_\mu \bar{\nu}_e$ or $\bar{\nu}_\mu \nu_e$ structure
of the beam, ratios between observed types of events (coming from each neutrino
species) could be used to establish a signal. 
Also polarization of the muon beam
could be useful because it allows to change the neutrino spectra and therefore
the event ratios in a predictable way. Possible channels involving $\tau$ 
leptons in the final state might be investigated by topological and 
kinematical methods as used by CHORUS and NOMAD. Finally, it could even be 
possible to obtain information about $CP$ violation by comparing 
event numbers from different channels.\\ 
Up to now we ignored in this section 
effects of neutrino oscillations. An incoming 
\aen could oscillate into a \amun{} and create via the aforementioned 
processes a $(-++)$ signal.  The relevant 
oscillation parameters are now 
given by atmospheric ($\Delta m^2 \simeq 10^{-3}$ eV$^2$) and 
CHOOZ ($\sin^2 2 \theta \ls 0.2$) data. 
Integrating the CC cross section 
of the \aen over a two--flavor formula and taking also into account 
the factor $10^{-4}$, yields numbers smaller than the ratio of process 
(\ref{process}) with the (CC + NC) cross section 
by at least one order of magnitude, even for a $L = 1$ km 
and $E_\mu = 50$ GeV option of the experiment.\\
For the other final states there is no SM background: Typical events with 
additional leptons are production of gauge bosons, which however are always 
accompanied with neutrinos or extra jets and thus in principle 
distinguishable.

\subsection{\label{obs?}Is it observable?}
Unfortunately, the bounds on neutrinos and their mixing severely restrict 
the prospects of detecting a signal from $B-L$ violating mass terms at 
the discussed experiments: For example, let us consider 
a 4 TeV muon source (be it a collider or just a storage ring) and 
the $\mu \mu$ channel. For the moment, we stick to one $m_i$. 
The maximum cross section is achieved for a 
mass eigenstate of about 10 GeV, 
$\sigma_{\mu\mu} (m_i \simeq 10 \; \rm GeV) \simeq 10^{-20}$ b. 
A few years of running with 
$\mbox{$\cal{L}$}$ $\simeq 10^{39}$ cm$^{-2}$ s$^{-1}$ per year could 
establish an observation. However, for the minimal allowed mass of 
84.1 GeV, the cross section reduces to 
$\sigma_{\mu\mu} (m_i = 84.1 \; \rm GeV) \simeq 2.0 \cdot 10^{-21}$ b, 
which is then further suppressed by the $U_{\mu i}$ limit 
to  $7.3 \cdot 10^{-26}$ b. Roughly the same number holds for the 
$ee$ channel, and for the $e\mu$ channel about three times this number. 
However, now the value from Eq.\ (\ref{imeffres}) comes into play: 
Assuming one mass eigenstate of 
81.8 GeV one gets $|U_{ei}|^2 < 9 \cdot 10^{-7}$, 
resulting in $\sigma_{ee} (m_i = 81.8 \; \rm GeV) \simeq 10^{-33}$ b! 
The cross section stays constant till $m_i = 6 \cdot 10^{5}$ GeV 
and scales with $m_i^{-2}$ 
for larger masses. For the $e \mu$ and $e\tau$ 
channel the cross sections are 
$\sigma_{e \mu} \simeq 2 \cdot 10^{-31} \; {\rm b}/m_i$[GeV] and 
$\sigma_{e \tau} \simeq 8  \cdot 10^{-31} \; {\rm b}/m_i$[GeV], 
respectively. 
Thus, the electron final states of process (\ref{process}) 
provide no real chance for observation. 
The importance of 
using the heavy neutrino bound from \obb{} 
was also discussed in \cite{heusch,inv2beta}.\\ 
Now we investigate possible event numbers: 
to be independent on the concrete values of the experimental parameters 
we calculate the charged and neutral current cross section by 
integrating over the energy spectrum (\ref{spectrum}) with the  
GRV 92 and 98 \cite{grv} parton distributions including 
$c$ and $b$ quark contributions. With this we give the maximal ratio 
(i.e.\ applying all limits of Eq.\ (\ref{neulim}) for 
the cross section) 
of the process (\ref{process}) as shown in Table \ref{table2}. 
We considered only the muonic and tauonic final states and 
took for the $\mu\tau$ channel the value $m_i$ = 84.1 GeV\@.
The last column 
gives the number of \mun (CC + NC) events with the parameter set given 
after Eq.\ (\ref{lumi}). 
Only the highest discussed energy provides a chance for observation. 
However, the realization of this kind of machine remains doubtful, 
but might be realized in a form of a new high--energy physics 
laboratory \cite{kingpev}. In Ref.\ \cite{inv2beta} 
it has been shown that the bounds from \obb{} on heavy
Majorana neutrinos can be evaded if one assumes rather baroque models,
which seem highly unnatural and are not considered here.\\
Though the numbers are no reason to be overoptimistic, 
the same final state we discussed might have contributions 
from other channels as the ones plotted in Fig.\ \ref{emu50R}. 
For \obb{} many limits on beyond--SM parameters 
were derived, see \cite{klap} for a review. 
A simple estimation shows the power of such a neutrino factory: 
For a 4 TeV energy and a 100 keV neutrino, the cross section is 
about $10^{-29}$ b. Other contributions might not need a helicity flip and 
are thus larger by roughly 
a factor of $(m_\nu/E_m)^2 \simeq 10^{13}$, where 
$E_m$ is the energy of the Majorana neutrino. With the mentioned 
$10^{39} $ cm$^{-2}$ s$^{-1}$ luminosity we would have $10^6$ events per year, 
a ``new--physics factory''.

\subsection{Outlook}
Will the situation change with future improved mass and mixing limits? 
First, the high number of neutrino interactions might have impact 
on the global fits for the mixing matrix elements. However, the limits 
cannot be expected to be improved by factors larger 
then $\cal{O}$(1). 
The LEP bound on the neutrino mass from \cite{L3} 
correspond to about 40 $\%$ of the used 
center of mass energy of 189 GeV\@, for simplicity we can assume that this 
will hold also for the upgrade energies. 
As other machines are concerned, 
at LHC \cite{LHC1,LHC2} or HERA \cite{HERA} investigation of masses of 
a few 100 GeV might be possible, but mostly only the electron channels were 
considered. Applying the \obb{} limit on heavy neutrinos (which has not 
been done in the analyses) 
on those results reduces the mass limits in those works. 
For the muon channels the results are significantly lower \cite{LHC2} 
or, as for HERA, not yet discussed. 
Regarding NLC, the pair production 
$e^+ e^- \to NN$ produces too small event numbers \cite{cvetic1}. 
Recently it was shown in 
\cite{alm2} that for $\sqrt{s} \gs 500$ GeV the ``indirect'' process  
$e^+ e^- \to \nu e^{\pm} W^{\mp}$ might probe Majorana 
masses up to the center of mass energy. The same holds for the 
$e\mu$ option of future colliders via  
$e^\pm \mu^\mp \to \nu l^\pm W^\mp$ \cite{alm3} and also Majorana neutrino 
pair production will be observable \cite{cvetic2}. 
Processes such as $e^- \gamma \ra \nu_e \alpha^- \beta^- W^+$ \cite{inv2beta} 
with $\alpha, \, \beta = \mu, \, \tau$ or 
$e^-e^- \ra \mu^- \mu^-$ \cite{pham} might also evade the \obb{} 
constraint but require center of mass energies in the 
same region as the ones discussed here.

\section{\label{funf}Conclusions}
To conclude, we did a full analysis of the analogue of \onbb{} at a 
neutrino factory whilst applying several experimental limits. 
The use of a detector right at the muon storage ring 
provides a very large number of neutrino interactions and 
for current and future mass limits the signals are perhaps 
observable at very high muon 
energies. Furthermore, if observed at lower energies, it is 
important to know how heavy Majorana neutrinos with SM   
couplings contribute to the events. 
The cross sections are very small but at least 
unaffected by oscillation phenomena. The limits on the 
effective Majorana mass matrix can be 
pushed down to physical values even from energies of $E_\mu = 500$ GeV on. 
For its ``inverse'' however, energies higher by a factor of more than 20 
are required. 
Signatures of the discussed events might be the only chance to 
find out about Majorana mass terms since most other related  
$B-L$ violating processes suffer from tiny ratios to the respective 
Standard Model events.  
Information on future limits on Majorana masses is 
mostly found in works concentrating on  
electron final states and thus unimportant when one incorporates 
\obb{} bounds on heavy Majorana neutrinos. Anyway, our results should 
not change dramatically even for new LEP limits on direct 
production or modified global fits. 
Contributions to the same final state 
without intermediate 
Majorana neutrinos (e.g.\ SUSY particles) 
however are a very realistic 
candidate for observation and the prospects for this will be addressed 
in future works. The process considered in this paper is then the 
relevant background signal.

\hspace{3cm}
\begin{center}
{\bf \large Acknowledgments}
\end{center}
This work has been supported in part (W.R.) by the
``Bundesministerium f\"ur Bildung, Wissenschaft, Forschung und
Technologie'', Bonn under contract No. 05HT9PEA5.
Financial support from the Graduate College
``Erzeugung und Zerf$\ddot{\rm a}$lle von Elementarteilchen''
at Dortmund university (W.R.) is gratefully acknowledged.

\newpage
\begin{table}[ht]
\begin{center}
\begin{tabular}{|c|c|c|c|c|c|c|} \hline
 $E_\mu$ & \mmm & \mmt & \mtta & \immm & \immt & \imtt \\ \hline
 50 &  25.2 &   57.6 & $1.2 \cdot 10^3 $ & 12.4 &  32.6 & 199.5  \\ \hline
 100 & 12.9 & 21.9 & 128.2 & 3.1 &   1.7 & 13.4 \\ \hline
 200 &  6.6 &  8.1 & 26.9 & 0.8 & 1.1 & 1.7 \\ \hline
 300 &  4.6 &  4.7 &  14.2 &   0.3 & 0.4 &  0.7 \\ \hline
 400 &  3.4 & 3.3 &  8.1 & 0.2 & 0.2 & 0.3 \\ \hline
 500 &  2.8 & 2.6 &  5.8 & 0.2 & 0.2 & 0.2  \\ \hline
 $10^3$ & 2.0 & 1.5 & 3.2  & $3.7 \cdot 10^{-2}$ & $3.7 \cdot 10^{-2}$      
& 0.1  \\ \hline
 $2 \cdot 10^3$ & 0.7 & 0.5 &   3.1 & $7.4 \cdot 10^{-3} $ 
& $7.4 \cdot 10^{-3} $ &$7.4 \cdot 10^{-3} $  \\ \hline
 $4 \cdot 10^3$ & 0.4 & 0.3 & 0.2 & $2.6 \cdot 10^{-3}$ & 
 $2.3 \cdot 10^{-3}$ & $2.8 \cdot 10^{-3}$ \\ \hline
 $10^4$ & 0.2 & 0.1 & 0.2 & $5.3 \cdot 10^{-4}$ & 
 $4.9 \cdot 10^{-4}$ &  $5.8 \cdot 10^{-4}$ \\ \hline
 $10^5$ &  $2.9 \cdot 10^{-2}$ & $1.4 \cdot 10^{-2}$ 
&  $1.4 \cdot 10^{-2}$ & $2.2 \cdot 10^{-5}$ & 
$1.6 \cdot 10^{-5}$ &  $1.6 \cdot 10^{-5}$ \\ \hline
 $10^6$ & $4.6 \cdot 10^{-4}$ & $2.9 \cdot 10^{-4}$ & 
$3.2 \cdot 10^{-4}$ & $1.4 \cdot 10^{-6}$ & $7.0 \cdot 10^{-7}$ & 
$1.4 \cdot 10^{-6}$ \\ \hline
\end{tabular}
\caption{\label{table1}Obtainable limits for \mab{} and \imab{} 
(in GeV and GeV$^{-1}$) for 
different muon energies in GeV\@. It holds 
$\mee{} \simeq \mmm \simeq 1/3 \; \meu{}$ 
and $\imee \simeq \immm{} \simeq 1/3$ \imem. For the number of 
events the parameter set given in Section \ref{paraset} is used. 
For \imab{} the limits from on $|U_{\alpha i}|^2$ from 
Eq.\ \ref{neulim} have been applied.}
\end{center}
\end{table}

\begin{table}[hb]
\begin{center}
\begin{tabular}{|c|c|c|c|c|} \hline
$E_\mu$ & $\frac{\D \sigma_{\mu\mu}}{\D \sigma^{CC + NC}}$ & 
$\frac{\D \sigma_{\mu\tau}}{\D \sigma^{\st CC + NC}}$ & 
$\frac{\D \sigma_{\tau\tau}}{\D \sigma^{\st CC + NC}}$ & $N^{CC + NC}$
\\ \hline \hline 
$ 50 $ & $ 1 \cdot 10^{-20} $ & $ 7 \cdot 10^{-21} $ 
& $ 7 \cdot 10^{-22} $ & $4 \cdot 10^{9}$ \\ \hline
$100 $ & $ 1 \cdot 10^{-19} $ & $ 1 \cdot 10^{-19} $ 
& $ 7 \cdot 10^{-20} $ & $8 \cdot 10^{9}$ \\ \hline
$200 $ & $ 7 \cdot 10^{-19} $ & $ 2 \cdot 10^{-18 }$ 
& $ 2 \cdot 10^{-18} $  & $2 \cdot 10^{10}$ \\ \hline
$300 $ & $ 3 \cdot 10^{-18} $ & $ 7  \cdot 10^{-18} $ 
& $ 1 \cdot 10^{-17} $  & $2 \cdot 10^{10}$ \\ \hline
$400 $ & $ 7 \cdot 10^{-18} $ & $ 2 \cdot 10^{-17} $ 
& $ 4 \cdot 10^{-17} $  & $4 \cdot 10^{10}$ \\ \hline
$500 $ & $ 1 \cdot 10^{-17} $ & $ 4 \cdot 10^{-17} $ 
& $ 7 \cdot 10^{-17} $  & $4 \cdot 10^{10}$ \\ \hline
$10^3 $ & $ 7 \cdot 10^{-17} $ & $ 4 \cdot 10^{-16} $ 
& $ 7 \cdot 10^{-16} $  & $8 \cdot 10^{10}$ \\ \hline
$ 2 \cdot 10^3 $ & $ 7 \cdot 10^{-16} $ & $ 3 \cdot 10^{-15} $ 
& $ 7 \cdot 10^{-15} $  & $1 \cdot 10^{11}$ \\ \hline
$4 \cdot 10^3 $ & $ 5 \cdot 10^{-15} $ & $ 2 \cdot 10^{-14} $ 
& $ 5 \cdot 10^{-14} $  & $2 \cdot 10^{11}$ \\ \hline
$10^4$ & $ 4 \cdot 10^{-14} $ & $ 2 \cdot 10^{-13} $ 
& $ 5 \cdot 10^{-13} $  & $5 \cdot 10^{11}$ \\ \hline
$10^5$  & $ 2 \cdot 10^{-12} $ & $ 1 \cdot 10^{-11} $ 
& $ 2 \cdot 10^{-11} $  & $3 \cdot 10^{12}$ \\ \hline
$10^6$  & $ 4 \cdot 10^{-11} $ & $ 8 \cdot 10^{-10} $ 
& $ 4 \cdot 10^{-10} $  & $8 \cdot 10^{12}$ \\ \hline
\end{tabular}
\caption{\label{table2}Maximal 
ratio of $\mun \! N \to \mu^- \alpha^+ \beta^+ X$ and 
sum of CC and NC for \mun{} for different final states and muon energies 
in GeV\@. Indirect bounds on mixing matrix elements are applied. The last 
column displays the expected number of (CC + NC) events from \mun{} 
with the optimistic parameter set given in Section \ref{paraset}.}
\end{center}
\end{table}

\clearpage
\newpage

\setlength{\unitlength}{1cm} 

\end{document}